\documentclass[aps,prd,twocolumn,groupedaddress]{revtex4-1}
\usepackage{graphicx}
\include{amssym}

\begin{document}
\title{Account for axial vector mesons in the $\eta\to\pi^+\pi^-\gamma$ and $\eta' \to\pi^+\pi^-\gamma$ decays}

\author{A. A. Osipov$^{1}$}
\email[]{aaosipov@jinr.ru}

\author{A. A. Pivovarov$^{1}$}
\email[]{tex$_$k@mail.ru}

\author{M. K. Volkov$^{1}$}
\email[]{volkov@theor.jinr.ru}

\author{M. M. Khalifa$^{2,3}$}
\email[]{mkhalifa@phystech.edu}

\affiliation{$^1$Bogoliubov Laboratory of Theoretical Physics, Joint Institute for Nuclear Research, Dubna, 141980, Russia}
\affiliation{$^2$Moscow Institute of Physics and Technology, Dolgoprudny, Moscow Region 141701, Russia}
\affiliation{$^3$Department of Physics, Al-Azhar University, Cairo 11751, Egypt}

\begin{abstract} 
The rates and spectra of the anomalous $\eta\to\pi^+\pi^-\gamma$ and $\eta' \to\pi^+\pi^-\gamma$ decays are calculated. The approach is based on the effective meson Lagrangian obtained in the Nambu-Jona-Lasinio model with vector and axial-vector mesons by integrating out quark fields. The resulting action is affected by mixing between members of pseudoscalar $J^{PC}=0^{-+}$ and axial-vector $1^{++}$ nonets that violates some low-energy theorems. In this note we point out that a gauge covariant procedure to diagonalize this mixing allows for consistent description of the $\eta\to\pi^+\pi^-\gamma$ and $\eta' \to\pi^+\pi^-\gamma$ decays.  
\end{abstract}

\maketitle

%%%%%%%%%%%%%%      Section 1 
\section{Introduction}
It is known that effective meson Lagrangians describing low-energy interactions of spin-0 and spin-1 states usually require a procedure for redefining the axial-vector field which is associated with spontaneous breaking of chiral symmetry \cite{Gasiorovicz69,Bando88,Meissner88}. By means of this procedure one eliminates the mixing between pseudoscalar (P) and axial-vector (A) fields (hereinafter for brevity we use the term "PA mixing"). In the Nambu-Jona-Lasinio (NJL) model, the PA mixing is induced by a one-quark-loop diagram connecting bound $\bar qq$ mesonic states in leading order in $1/N_c$ \cite{Osipov85,Volkov86,Ebert86}. The corresponding contribution is proportional to the constituent quark mass, and therefore is generated dynamically through a partial Higgs mechanism. In the world of zero bare mass for the up, down and strange quarks, this does not break the original $U(3)$ flavor symmetry. However, in the real world, with small but non-zero bare quark masses, this contribution violates both the $U(3)$ nonet symmetry and the $SU(3)$ flavor symmetry. Thus the PA mixing is an additional source of flavor symmetry breaking in the effective meson Lagrangian \cite{Morais17}. 

The purpose of this paper is to demonstrate that PA mixing affects a non-resonant contribution in the anomalous $\eta / \eta'\to\pi^+\pi^-\gamma$ decays of eta mesons, i.e. the box anomaly. The selection of processes is not accidental. These decays are reasonably well studied experimentally and make it possible to measure the magnitude of the non-resonant contribution \cite{Abele97}; moreover, they are sensitive to the flavor symmetry breaking. These modes have been also extensively investigated in the framework of the chiral perturbation theory (ChPT) \cite{Bijnens90}, and in different models based on specific chiral Lagrangians with vector mesons \cite{Schechter84,Pak85a,Picciotto92,Kuraev92,Kuraev95,David03,David10}. Despite the great work done, a violation of flavor symmetry through the mechanism of eliminating PA mixing has not yet been addressed in the literature. This can be partly explained by the problem that arises when considering axial-vector mesons. After elimination of the PA mixing, some meson amplitudes receive additional contributions that violate a number of low-energy theorems of current algebra and PCAC (partially conserved axial-vector current) hypothesis \cite{Pak85,Wakamatsu89}. Owing to this problem, accounting for contributions from axial-vector mesons to the $\eta / \eta' \to\pi^+\pi^-\gamma$ amplitudes is not a straightforward issue. 

Early attempts to solve the problem were based on the naive subtraction of vertices of the effective photon-meson Lagrangian, which violate the low-energy theorems \cite{Pak85}. This was done by assuming that in the low-energy region the vector-meson dominance (VMD) hypothesis holds exactly. The latter assumption can hardly be justified from QCD. In addition, there are reasonable grounds to believe that deviations from VMD can even occur when considering only pseudoscalar and vector mesons, i.e. before the inclusion of axial-vector states to the effective action \cite{Fujiwara85}. Note that VMD-based subtractions do not account for residual effects associated with an explicit violation of chiral symmetry, and therefore do not meet the purpose pursued by us here.

Apart from the language of VMD-based subtractions, there is a more practical way to discuss the problem \cite{Osipov18a,Osipov18b,Osipov20}. It is an approach based on a QCD inspired effective NJL Lagrangian \cite{Ebert83,r14,Ebert86}. In accord with a general assertion of QCD that meson physics in the large $N_c$ limit is described by the tree diagrams of an effective local Lagrangian \cite{Hooft74a,Witten79}, the NJL model associates with any mesonic vertex the local part of the underlying quark loop diagram. In this quark-loop based approach, contributions ensuring the fulfilment of low-energy theorems in the presence of axial-vector mesons are generated by the PA mixing elimination procedure itself. An important role is given to fermionic triangle-loop graphs which are (superficially) linearly divergent. Owing to the linear divergence, shifting the integration momentum in the closed loop changes the value of the integral, so that there is an essential ambiguity which can be used to ensure Ward identities \cite{Bell69,Jackiw72,Jackiw00}.  

This idea has been applied recently to show how surface terms of some anomalous triangle Feynman diagrams can be used to satisfy Ward identities for $a_1\to\pi^+\pi^-\gamma$ and $\gamma\to 3\pi$ decays \cite{Osipov18a,Osipov18b,Osipov20}. It is important that necessary triangle diagrams arise only due to gauge-covariant diagonalization of $\pi a_1$ mixing. A naive (non gauge-covariant) diagonalization of the $\pi a_1$ mixing for both processes leads to a contradiction with the corresponding low-energy theorems. 

What is special about the gauge-covariant diagonalization? Let us remind that the standard diagonalization procedure consists in redefining the axial-vector field $a_\mu$ as
\begin{equation}
\label{ngcr}
a_\mu = a_\mu^{(\mbox{\scriptsize phys})} + \frac{\partial_\mu \pi}{a g_\rho f_\pi}, 
\end{equation}
where $a_\mu =\tau_i a^i_\mu$, $\pi =\tau_i \pi^i$, $\tau_i$ are hermitian $SU(2)$ Pauli matrices, $i=1,2,3$; $f_\pi \simeq 93\,\mbox{MeV}$, the coupling constant $g_\rho \simeq \sqrt{12\pi}$ is fixed by relating it to the  $\rho\to\pi\pi$ decay width; $a$ is a dimensionless parameter which comes out as a result of diagonalization   
\begin{equation}
\label{a}
\frac{1}{a} = \frac{g_\rho^2 f_\pi^2}{m_\rho^2},
\end{equation}
where $m_\rho$ is the mass of the $\rho$ meson. 

In the presence of electromagnetic interactions the replacement (\ref{ngcr}) is not a $U(1)_{\mbox{\scriptsize em}}$ gauge-covariant one. Our alternative diagonalization procedure is suggested by the appearance of the $U(1)_{\mbox{\scriptsize em}}$ gauge-covariant derivative $\mathcal D_\mu \pi$ instead of the non gauge-covariant one $\partial_\mu \pi$ in (\ref{ngcr})
\begin{equation}
\label{cov}
a_\mu = a_\mu^{(\mbox{\scriptsize phys})}  + \frac{\mathcal D_\mu \pi}{ag_\rho f_\pi}, \quad \mathcal D_\mu \pi =\partial_\mu \pi -ieA_\mu [Q,\pi ],
\end{equation}
where $Q=(1+3\tau_3)/6$ is the electric charge matrix of quark fields, $e=\sqrt{4\pi\alpha}$ is the positron charge. Notice that the coupling of the electromagnetic field $A_\mu$ to pions (and as a consequence to quarks in the form $\bar q\gamma^\mu\gamma_5 \mathcal D_\mu \pi q$) can be carried out unambiguously using the gauge principle. 

\begin{figure}
\resizebox{0.25\textwidth}{!}{%
  \includegraphics{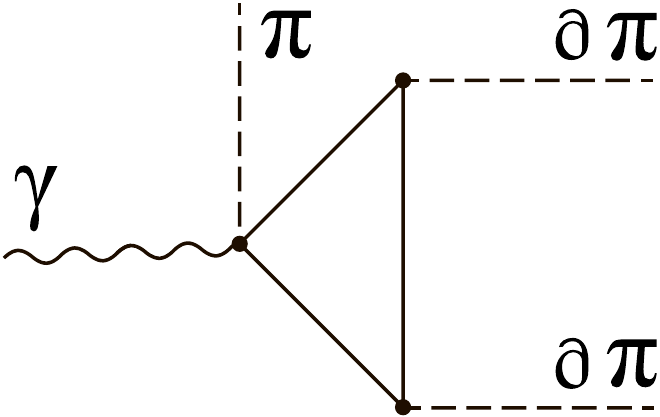}
}
\caption{The fermionic one-loop graph contributing to the low-energy $\gamma\pi\pi\pi$ amplitude if the surface term is not ignored. It is suggested that such diagrams can be used to satisfy the Ward identities violated by the $\pi a_1$ mixing.}
\label{fig1}      
\end{figure}

One might think of criticizing new replacement (\ref{cov}) on the grounds that in accord with the known equivalence theorem in the axiomatic field theory (Haag's theorem \cite{Haag58}), as well as in its Lagrangian version (Chisholm's theorem \cite{Chisholm61,Salam61}), both replacements (\ref{ngcr}) and (\ref{cov}) are equivalent and therefore should lead to the same physical content of the theory. Indeed, most likely this is true for the real part of the effective action \cite{Osipov18b,Osipov18c,Osipov19} and this probably would be true for the anomalous (imaginary) part of the action if we, as usual, neglected the contribution of the surface terms. It should be noted, however, that the replacement (\ref{cov}) is a source of new anomalous triangle diagrams with photons possessing surface terms that cannot be ignored because they are important to fulfil the requirements of Ward identities. For instance, in Fig.~\ref{fig1} we show a Feynman diagram which is zero if one neglects the contribution of the surface term. It is this diagram that allows us to ensure the fulfilment of Ward identities for the anomalous $\gamma\pi\pi\pi$ vertex \cite{Osipov20}.

%%%%%%%%%%%%%%%    FIG-1    %%%%% 

The paper is organized as follows. In Sec. II we discuss the reason for mixing in the system of $\eta$ and $\eta'$ mesons. We recall the Witten-Veneziano approach to resolve the $U(1)$ problem and show that this framework leads naturally to the $\eta-\eta'$ mixing angle $\theta_P\simeq -20^\circ$. In Sec. III we discuss the $\gamma\gamma$ widths of the $\eta$ and $\eta'$. We define here the notation and use the experimental data on these modes to fix the main parameters involved in the description of the $\eta$ and $\eta'$ system. Radiative $\eta / \eta'\to\pi^+\pi^-\gamma$ decays are discussed in Sec. IV. In this section we show that the axial-vector mesons through the $PA$ mixing mechanism affect the box anomaly at lowest order of flavor symmetry breaking. That leads to the difference in the description of flavor symmetry breaking effects for box and triangle anomalies. This section contains the main result of the paper. Our conclusions are found in Sec. V.

%%%%%%%%%%%%%%      Section 2 
\section{$\eta -\eta'$ mixing}
In the limit of zero bare mass for the up, down and strange quarks $\hat m_u=\hat m_d=\hat m_s=0$, the pseudoscalar states belonging to the $SU(3)$ octet become massless Goldstone bosons, but the ninth pseudoscalar state, the $SU(3)$ singlet $\eta_0$, remains massive $m_{\eta_0}^2=3\lambda_\eta^2/N_c$ due to the $U(1)$ anomaly \cite{Witten79b}. With the explicit breaking of chiral symmetry $\hat m_u=\hat m_d\ll \hat m_s$ (isospin is assumed to be exact) the octet pseudoscalar masses become nonzero and are related, at first order in the quark mass expansion, by the Gell-Mann-Okubo formula $3m^2_{\eta_8}=4m_K^2-m_\pi^2$ \cite{Gell-Mann61,Gell-Mann62,Okubo62}, where $\eta_8$ is the eighth member of the $SU(3)$ octet. 

In lowest order of chiral expansion, if $\hat m_u=\hat m_d\neq \hat m_s$, the massive $SU(3)$ singlet $\eta_0$ mixes with $\eta_8$ producing the mass matrix $m^2_{ab}$ ($a,b=8,0$) which in the $\eta_8$, $\eta_0$ basis can be written as follows \cite{Veneziano79}
\begin{equation}
\label{MM}
\left(
\begin{array}{cc} \frac{4}{3}m_K^2-\frac{1}{3}m^2_\pi   &-\frac{2}{3}\sqrt 2(m_K^2-m_\pi^2) \\ -\frac{2}{3}\sqrt 2(m_K^2-m_\pi^2) & \frac{2}{3}m_K^2+\frac{1}{3}m^2_\pi+\frac{3\lambda_\eta^2}{N_c}
\end{array}
\right).
\end{equation}
Taking the trace of this matrix one obtains the Veneziano formula 
\begin{equation}
m^2_{88}+m^2_{00}=m^2_{\eta}+m^2_{\eta'}=2m_K^2+\lambda_\eta^2, 
\end{equation}
with the phenomenological estimate $\lambda_\eta^2= 0.726\,\mbox{GeV}^2$.

The physical eigenstates $\eta, \eta'$ are given by
\begin{eqnarray}
\label{Mix}
&&\eta =\cos\theta_P\, \eta_8 - \sin\theta_P\, \eta_0 \nonumber \\
&&\eta' =\sin\theta_P\, \eta_8 + \cos\theta_P\, \eta_0,
\end{eqnarray}  
where $\theta_P$ is a mixing angle. Diagonalization of (\ref{MM}) then yields
$$ %\begin{equation}
m_{\eta', \eta}^2=m_K^2+\frac{3\lambda_\eta^2}{2N_c}\pm \sqrt{\left(m_K^2-m_\pi^2-\frac{\lambda_\eta^2}{2N_c}\right)^2+2\frac{\lambda_\eta^4}{N_c^2}}, 
$$ %\end{equation}
with the following restriction for $\theta_P$ 
\begin{equation}
\label{angle}
\tan 2\theta_P=-\frac{\frac{4}{3}\sqrt 2 (m_K^2-m_\pi^2)}{\lambda_\eta^2 -\frac{2}{3}(m_K^2-m_\pi^2)}.
\end{equation}
The mass matrix (\ref{MM}) leads automatically to Schwinger's mass relation \cite{Schwinger64} for the nonet of pseudoscalar mesons
\begin{equation}
\frac{\left(m_{\eta'}^2-m_\pi^2\right)\left(m_{\eta}^2-m_\pi^2\right)}{\left(m_{\eta'}^2+m_\eta^2-2m_K^2\right)}=\frac{4}{3} \left(m_{K}^2-m_\pi^2\right). 
\end{equation}
The experimental values for left and right sides of this formula are $0.35\,\mbox{GeV}^2$ and $0.30\,\mbox{GeV}^2$ correspondingly. This is a quite good result for the approximation used. 

The other consequence of this consideration is the value of the pseudoscalar mixing angle $\theta_P\simeq -18.3^\circ$. This lowest order analysis does not include the leading logarithmic corrections arising from the meson one-loop diagrams. Nonetheless, the estimate obtained is quite compatable with that given in the full one-loop analysis of the ChPT: $\theta_P =-20^\circ\pm 4^\circ $ \cite{Gasser85}. 

Since we are interested in consistency with the simplest possible situation, the old parameterization \cite{Chanowitz75} in terms of two (octet $f_8$ and singlet $f_0$) decay constants and one $\eta-\eta'$ mixing angle $\theta_P$ will be used through the whole paper. This is sufficient both to describe the $\eta, \eta'$ radiative decay results considered here, and to demonstrate the main idea of our approach -- the effect of $PA$-mixing on the box anomaly. This does not exclude the further extension of the idea to more involved parameterizations in terms of either two octet-singlet decay constants and two mixing angles \cite{Leutwyler98}; or in terms of strange and non-strange decay constants and only one mixing angle, see e.g. Refs. \cite{Schechter93,Kroll98,Kroll99,Feldmann00,Osipov07}; or even in the form used within the context of the hidden local symmetry \cite{David03}.

%%%%%%%%%%%%%%      Section 3 
\section{$\eta\to\gamma\gamma$ and $\eta'\to\gamma\gamma$ decays}
The dominant two-photon decay modes of $\eta$ and $\eta'$ are described by matrix elements which do not preserve the intrinsic parity. Such anomalous interactions were treated by Wess and Zumino \cite{Wess71}. The topological content of the anomalous action was clarified by Witten \cite{Witten83}. The corresponding piece of the Wess-Zumino Lagrangian is given by 
\begin{equation}
\label{Pgg}
\mathcal{L}_{\phi\gamma\gamma}=-\frac{3}{2}F^\pi e^{\mu\nu\alpha\beta} \partial_\mu A_\nu \partial_\alpha A_\beta \,\mbox{tr}(Q^2 \phi ).
\end{equation}
Here the factor $F^\pi$ is given by 
\begin{equation}
F^{\pi}=\frac{N_c e^2}{12\pi^2 f_\pi}=0.025\,\mbox{GeV}^{-1}.
\end{equation}
The nonet of the pseudoscalar fields is described by the matrix $\phi =\sum_{i=0}^{8}\phi_i \lambda_i$; matrices acting in flavour space, $\lambda_i$, are the standard $SU(3)$ Gell-Mann matrices for $i\neq 0$, and $\lambda_0=\sqrt \frac{2}{3}$. These matrices obey the following basic trace properties: $\mbox{tr}\lambda_i=\sqrt 6 \delta_{i0}$, $\mbox{tr}(\lambda_i\lambda_j)=2\delta_{ij}$. As a result, we have
\begin{eqnarray}
\mbox{tr}(Q^2 \phi )=\frac{1}{3}\left[\pi^0\right. &\!+\!&\frac{\eta}{\sqrt 3}\left(\cos\theta_P-2\sqrt 2 \sin\theta_P\right)  \nonumber \\
&\!+\!&\left. \frac{\eta'}{\sqrt 3}\left(\sin\theta_P+2\sqrt 2 \cos\theta_P\right)  \right].
\end{eqnarray}

%%%%%%%%%%%%%%%    FIG-2    %%%%% 
\begin{figure}
\resizebox{0.30\textwidth}{!}{%
  \includegraphics{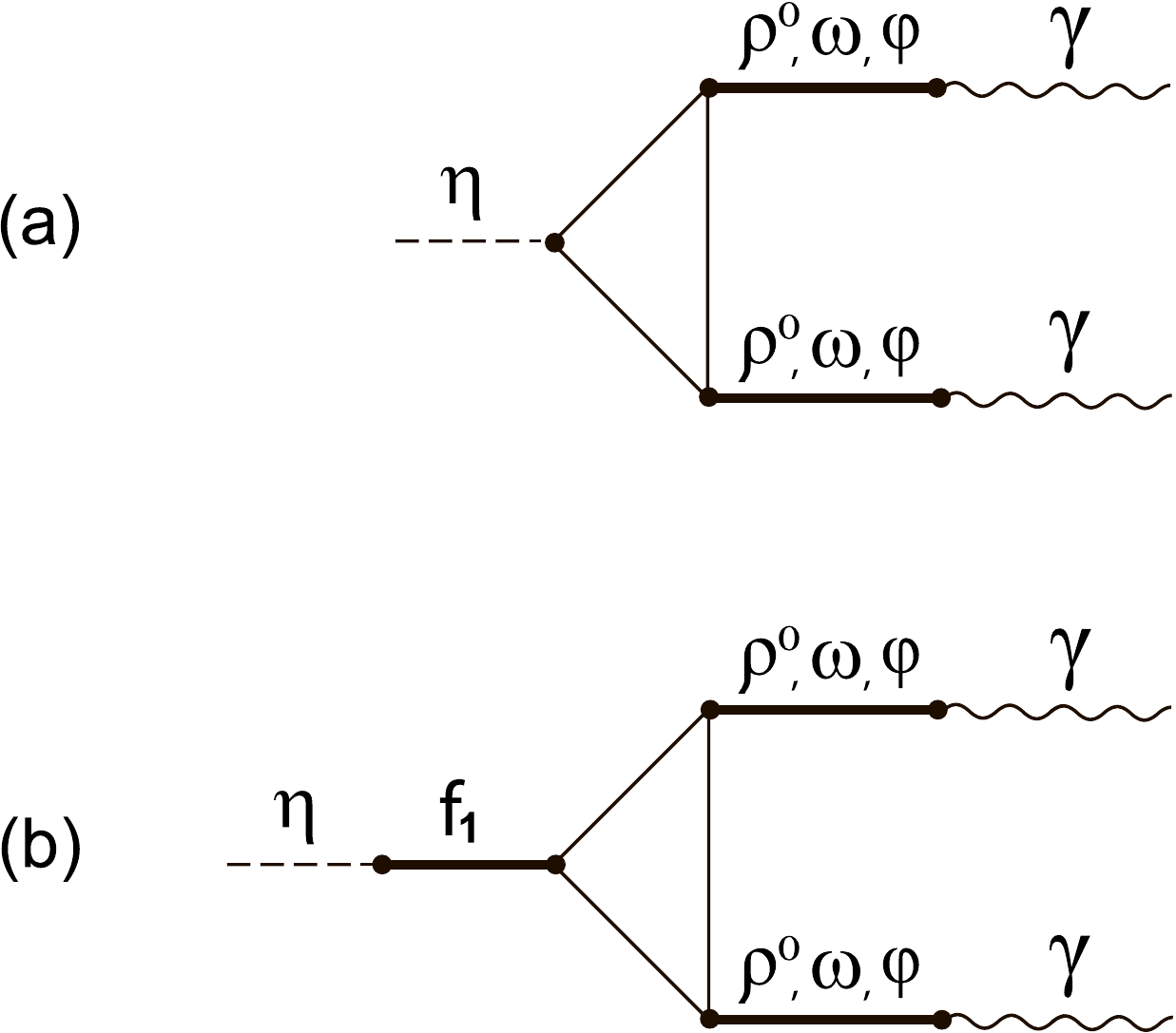}
}
\caption{Two graphs for the $\eta/\eta'\to\gamma\gamma$ decays in the NJL model with vector meson dominance. The $\eta f_1$ mixing induced graph (b) is forbidden due to the Landau-Yang theorem.}
\label{fig2}      
\end{figure}

The Lagrangian density (\ref{Pgg}) describes perfectly well the $\pi^0\to\gamma\gamma$ decay. However, in order to deal with interactions involving $\eta$ or $\eta'$ mesons, one should take into account the $SU(3)$ and nonet symmetry breaking effects. It is generally believed that these effects can be taken into account through a naive replacement of pseudoscalar decay constants \cite{Chanowitz75,Chanowitz80,Gilman87,Holstein02}. Namely, the $\eta/\eta'\to\gamma\gamma$ decays are usually described at the chiral point by the following amplitudes 
\begin{equation}
A_{\eta, \eta'\to\gamma\gamma}=F^{\eta, \eta'}(0)e^{\mu\nu\alpha\beta}\epsilon_\mu^* (q_1) q_{1\nu}\epsilon_\alpha^* (q_2) q_{2\beta},
\end{equation}
where couplings have the values 
\begin{eqnarray}
F^{\eta}(0)&=&\frac{F^\pi}{\sqrt 3} \left(\frac{f_\pi}{f_8}\cos\theta_P-2\sqrt 2 \frac{f_\pi}{f_0}\sin\theta_P\right),\\
F^{\eta'}(0)&=&\frac{F^\pi}{\sqrt 3}\left(\frac{f_\pi}{f_8}\sin\theta_P+2\sqrt 2 \frac{f_\pi}{f_0}\cos\theta_P\right)
\end{eqnarray}
The decay constants $f_8$ and $f_0$ are defined from axial-vector current expectation values: $\langle 0|J_\mu^{A8}|\eta_8\rangle =if_8 p_\mu$, $\langle 0|J_\mu^{A0}|\eta_0\rangle =if_0 p_\mu$; $\epsilon_\mu (q)$ is a photon polarization with 4-momentum $q_\mu$. Using the experimental numbers of the two-photon decay widths, and the ratio $f_8/f_\pi\simeq 1.3$ from the extended ChPT one can obtain that $\theta_P=-20^\circ$, and $f_0/f_\pi\simeq 1.04$. It is these parameter values that we will use in our subsequent numerical estimates.

The meson vertices (\ref{Pgg}) can be obtained through the direct calculation of the anomalous quark triangle  diagrams shown in Fig.\ref{fig2} by using, for instance, the NJL model with spin-1 mesons included. One can show that the diagram in Fig.\ref{fig2}b which is generated by $\eta -f_1$ transitions does not contribute. In other words, the $U(3)$ version of the shift (\ref{ngcr}) does not modify a result of quark loop calculations. The reasoning behind it is the Landau-Yang theorem \cite{Landau48,Yang56} which states that a massive unit spin particle cannot decay into two on shell massless photons. In particular, the theorem forbids the $f_1\to\gamma\gamma$ decays, where $f_1$ is a short hand for either $f_1(1285)$ or $f_1(1420)$ axial-vector states which can mix with the $\eta, \eta'$.

%%%%%%%%%%%%%%      Section 4
\section{$\eta /\eta'\to\pi^+\pi^-\gamma $ decays}
We can now confront the main subject of our paper -- that of the $\eta /\eta'\to\pi^+\pi^-\gamma$ decays. In the Wess-Zumino Lagrangian the piece responsible for these decays has a form %\begin{widetext}
\begin{eqnarray}
\label{gppp}
&&\mathcal{L}_{\gamma\phi\phi\phi}=\frac{iF^\pi}{2ef_\pi^2} e^{\mu\nu\alpha\beta} A_\mu \,\mbox{tr}(Q\partial_\nu \phi  \partial_\alpha \phi \partial_\beta \phi ) \nonumber \\
&&=\frac{iF^\pi}{ef_\pi^2} e^{\mu\nu\alpha\beta} A_\mu \left[\partial_\nu\pi^0+\frac{\partial_\nu\eta}{\sqrt 3}\left(\cos\theta_P-\sqrt 2\sin\theta_P\right) \right. \nonumber \\
&&+\left.\frac{\partial_\nu\eta'}{\sqrt 3}\left(\sin\theta_P+\sqrt 2\cos\theta_P\right)\right]\partial_\alpha\pi^+\partial_\beta\pi^- +\ldots
\end{eqnarray}

Again it is necessary to feed this Lagrangian density with effects of nonet and $SU(3)$ symmetry breaking when considering the $\eta, \eta'$ decays. The corresponding standard modifications will be introduced later.

%%%%%%%%%%%%%%%    FIG-3    %%%%% 
\begin{figure}
\resizebox{0.35\textwidth}{!}{%
  \includegraphics{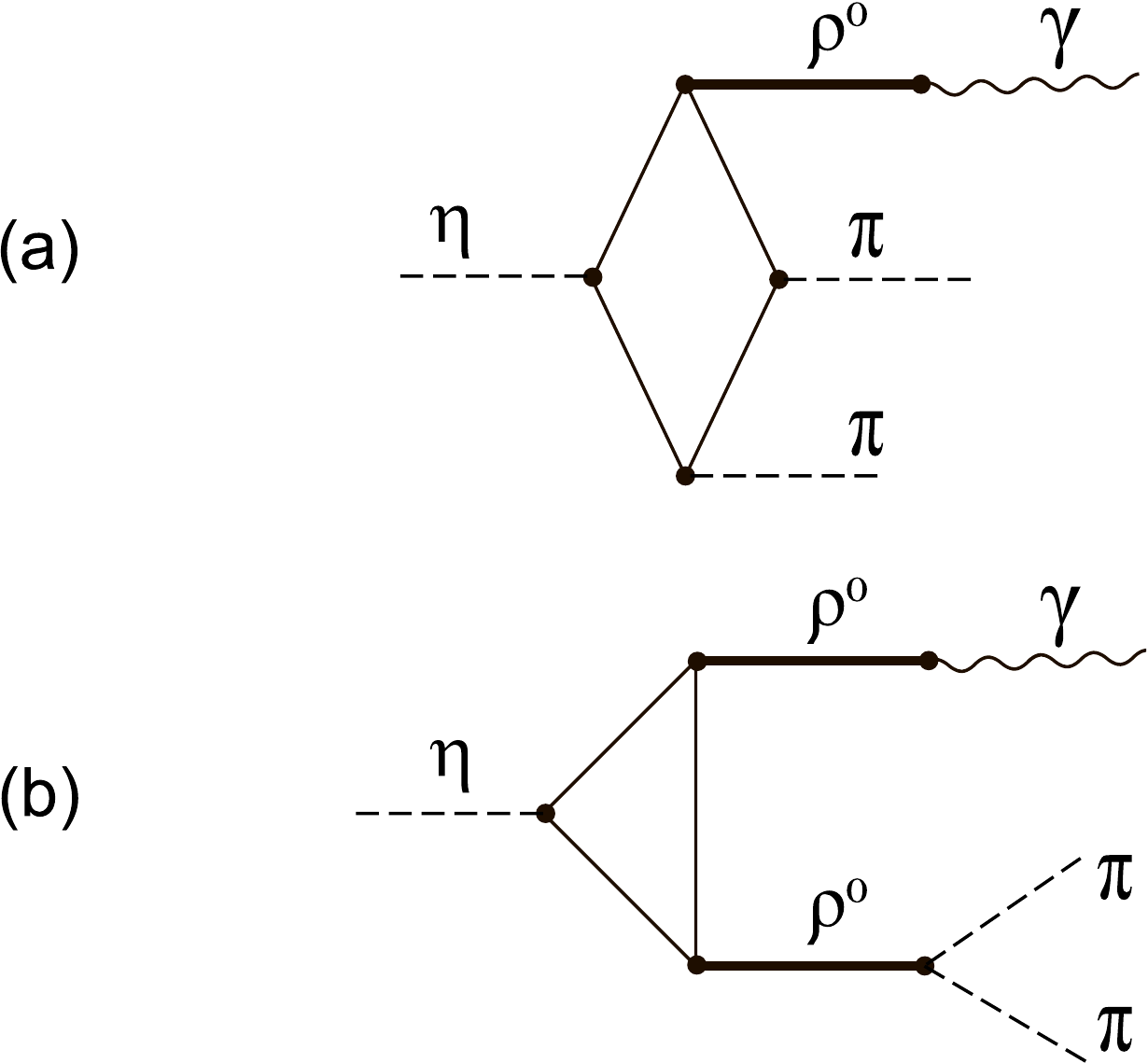}
}
\caption{Two graphs for the $\eta/\eta'\to\pi^+\pi^-\gamma$ decays in the NJL model with vector meson dominance. The box graph (a) is affected by the $PA$ mixing effects. On the contrary, the triangle anomaly $\eta\gamma\rho$ (and $\eta'\gamma\rho$) of the graph (b) is not affected by the $PA$ mixing.}
\label{fig3}      
\end{figure}

At the one-quark-loop level the $\eta\to\pi^+\pi^-\gamma$ amplitude receives contributions from the box and $\rho$-exchange diagrams, shown in Fig.\ref{fig3}. In the following we will refer to them as the box and the triangle anomalies. An essential difference between the box and the triangle diagrams is that the box suffers from effects induced by the shift of axial-vector fields (\ref{ngcr}). These shifts violate the low-energy theorems \cite{Pak85,Wakamatsu89} and need a special consideration. Indeed, the direct calculations of the box graphs of Fig.\ref{fig3}a give    

\begin{figure}
\resizebox{0.35\textwidth}{!}{%
\includegraphics{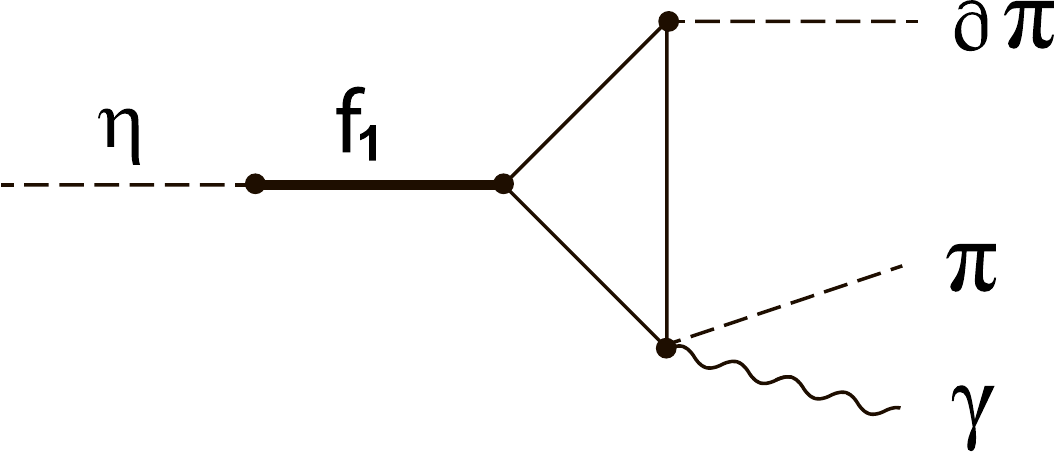}
}
\caption{The triangle quark-loop graph contributing to the $\eta/\eta'\to\pi^+\pi^-\gamma$ decay in the NJL model with the gauge-covariant $\pi a_1$ and $\eta f_1$ diagonalizations (\ref{cov}). }
\label{fig4}      
\end{figure}

\begin{widetext}
\begin{equation}
\label{box}
A_{\mbox{\scriptsize box}}=\frac{e N_c}{12 \pi^{2} f_{\pi}^{3}} \frac{1}{\sqrt 3} \left( \frac{f_\pi}{f_8}\cos\theta_P - \sqrt 2 \frac{f_\pi}{f_0} \sin\theta_P \right) \left[1 - \frac{2}{a} - \frac{1}{a_\eta}+\frac{1}{a}\left(\frac{2}{a_\eta}-\frac{1}{2a}\right) +\frac{1}{8a^2a_\eta} \right] e_{\mu\nu\alpha\beta}\, \epsilon^{*\mu}(p_{\gamma})\, p_{\gamma}^\nu p_{+}^\alpha p_{-}^\beta , 
\end{equation}
where $a$ is given by (\ref{a}). The parameter $a_\eta$ differs from $a$ only in that it arises from the diagonalization of $\eta f_1$ mixing. In the limit of exact $U(3)$ symmetry $a_\eta$ coincides with $a$. Our notations for 4-momenta of the photon and charged pions $p_{\gamma}^\mu, p_{+}^\mu$ and $p_{-}^\mu$ are obvious.  

In its turn, the $\rho$-exchange diagram shown in Fig. \ref{fig3}b leads to the amplitude	
\begin{eqnarray}
\label{rho}
A_{\rho}&=& \frac{e N_c}{4\pi^2f_\pi^3} \frac{1}{\sqrt 3} \left( \frac{f_\pi}{f_8}\cos\theta_P - \sqrt 2 \frac{f_\pi}{f_0} \sin\theta_P \right)\frac{g_{\rho}^{2} f_\pi^2}{m_{\rho}^{2} - q^{2}} e_{\mu\nu\alpha\beta}\, \epsilon^{*\mu}(p_{\gamma})\, p_{\gamma}^\nu p_{+}^\alpha p_{-}^\beta \nonumber \\
&=& \frac{e N_c}{12\pi^2f_\pi^3} \frac{1}{\sqrt 3} \left( \frac{f_\pi}{f_8}\cos\theta_P - \sqrt 2 \frac{f_\pi}{f_0} \sin\theta_P \right)\frac{3}{a} \left(1+\frac{q^2}{m_\rho^2-q^2}\right) e_{\mu\nu\alpha\beta}\, \epsilon^{*\mu}(p_{\gamma})\, p_{\gamma}^\nu p_{+}^\alpha p_{-}^\beta ,
\end{eqnarray}
where $q = p_{+} + p_{-}$. 

The sum of Eqs. (\ref{box}) and (\ref{rho}) is
\begin{eqnarray}
A_{\mbox{\scriptsize box}}+A_{\rho} &=& \frac{e N_c}{12 \pi^{2} f_{\pi}^{3}} \frac{1}{\sqrt 3} \left( \frac{f_\pi}{f_8}\cos\theta_P - \sqrt 2 \frac{f_\pi}{f_0} \sin\theta_P \right) e_{\mu\nu\alpha\beta}\, \epsilon^{*\mu}(p_{\gamma})\, p_{\gamma}^\nu p_{+}^\alpha p_{-}^\beta  \nonumber \\
&\times& \left[1 + \frac{1}{a} - \frac{1}{a_\eta}+\frac{1}{a}\left(\frac{2}{a_\eta}-\frac{1}{2a}\right) +\frac{1}{8a^2a_\eta} +\left(\frac{3}{a}\right)\frac{q^2}{m_\rho^2-q^2}\right].
\end{eqnarray}

This clearly shows that the expression in square brackets does not turn into unity (at $q^2=0$) even when the $U(3)$ symmetry is exact. The latter contradicts to the requirements of the low-energy theorem (\ref{gppp}). As we have already noted, the way out of this problem is to use the gauge-covariant diagonalization (\ref{cov}), which leads to the consideration of the additional diagram shown in Fig.\ref{fig4}. Taking into account the contribution of this diagram we obtain 
\begin{eqnarray}
A_{\eta\to\pi\pi\gamma}=A_{\mbox{\scriptsize box}}+A_{\rho} + A_{\mbox{\scriptsize new}}
&=&\frac{e N_c}{12 \pi^{2} f_{\pi}^{3}} \frac{1}{\sqrt 3} \left( \frac{f_\pi}{f_8}\cos\theta_P - \sqrt 2 \frac{f_\pi}{f_0} \sin\theta_P \right) e_{\mu\nu\alpha\beta}\, \epsilon^{*\mu}(p_{\gamma})\, p_{\gamma}^\nu p_{+}^\alpha p_{-}^\beta \nonumber \\
&\times& \left[1 + \frac{1}{a} - \frac{1}{a_\eta}+\frac{1}{a}\left(\frac{2}{a_\eta}-\frac{1}{2a}\right) +\frac{1-12 b}{8a^2a_\eta} +\left(\frac{3}{a}\right)\frac{q^2}{m_\rho^2-q^2}\right].
\end{eqnarray}
Owing to the shift ambiguity related to the formal linear divergence of the one-loop triangle integral, the result depends on the undetermined coupling $b$, which survives in the final expression \cite{Bell69,Jackiw72,Jackiw00,Baeta01}. Observing that 
\begin{equation}
\frac{1}{a} - \frac{1}{a_\eta}+\frac{1}{a}\left(\frac{2}{a_\eta}-\frac{1}{2a}\right) +\frac{1-12 b}{8a^2a_\eta}=\left(\frac{3}{2a^2}+\frac{1-12b}{8a^3} \right) +\left(\frac{1}{a}-\frac{1}{a_\eta}\right)\left(1-\frac{2}{a}-\frac{1-12b}{8a^2}\right),
\end{equation}
\end{widetext}
we see that the coupling $b$ can be uniquely fixed in accord with the low-energy theorem (\ref{gppp}), namely $b=a+1/12$. Thus, we finally obtain
\begin{eqnarray}
\label{tot1}
A_{\eta\to\pi\pi\gamma}&=& \frac{e N_c}{12 \pi^{2} f_{\pi}^{3}} \frac{1}{\sqrt 3} \left( \frac{f_\pi}{f_8}\cos\theta_P - \sqrt 2 \frac{f_\pi}{f_0} \sin\theta_P \right) \\
&\!\!\!\!\!\!\!\!\!\!\!\!\!\times&\!\!\!\!\!\!\!\!\left[1 +\delta +\left(\frac{3}{a}\right)\frac{q^2}{m_\rho^2-q^2}\right] e_{\mu\nu\alpha\beta}\, \epsilon^{*\mu}(p_{\gamma})\, p_{\gamma}^\nu p_{+}^\alpha p_{-}^\beta ,\nonumber 
\end{eqnarray}
where
\begin{equation}
\label{delta}
\delta=\left(\frac{1}{a}-\frac{1}{a_\eta}\right)\left(1-\frac{1}{2a}\right).
\end{equation}

%%%%%%%%%%%%   FIG. 4   %%%%%%%%

Some comments about formula (\ref{tot1}) are in order. Let us first notice that in the case of exact $U(3)$ symmetry, $\delta=0$. This follows from the method used to fix the constant $b$. The method is based on satisfying Ward identities for the amplitudes of the processes $\gamma\to\pi^0\pi^+\pi^-$ and $\eta\to\pi^+\pi^-\gamma$. In both cases this requirement yields the same value of $b$ (to compare see \cite{Osipov20}).

Next, among the implicit assumptions commonly used in describing $\eta /\eta' \to\pi^+\pi^-\gamma$ decays there is the hypothesis that $SU(3)$ and nonet symmetry breaking act in exactly the same way for the triangle and box anomalies. This would also be true in our approach, if $\delta$ would be zero. It should be noted however that the effects of $U(3)$ symmetry breaking in the triangle and box anomalies differ, if $\delta\neq 0$. In the case considered here, this difference is related to the inclusion of spin-1 mesons.

For our purpose we do not need to calculate $a_\eta$ explicitly. We are faced here with a simpler task -- to demonstrate the main consequence of using the diagram Fig.\ref{fig4} in solving the problem of $PA$ mixing in $\eta /\eta' \to\pi^+\pi^-\gamma$ decays. For that it is enough to know that $a_\eta\neq a$ when $U(3)$ symmetry is broken, i.e. that $\delta\neq 0$. The latter follows from the observation that $PA$ transition is proportional to the squared quark mass. The amplitude of $\pi a_1$ transition is described by the one-quark-loop graph containing only light $u$ and $d$ quarks. This gives $a$. The $\eta\to f_1$ amplitude contains also the strange quarks, and this is why $a_\eta\neq a$. The value of $\delta$ can be extracted from the experiment.

One should still include unitarity effects to the amplitude (\ref{tot1}) via final state interactions. This is very important for $\eta'\to\pi^+\pi^-\gamma$ decay, where the physical region is $4m_\pi^2\leq q^2\leq m_{\eta'}^2$, and less important for the $\eta\to\pi^+\pi^-\gamma$ decay, where the $\rho$-meson pole is out of the physical region. One very obvious approach is simply to include the (energy-dependent) width of the $\rho$-meson in the propagator in the vector-dominance form 
\begin{equation}
\frac{q^2}{m^2_\rho -q^2} \to \frac{q^2}{m_\rho^2-q^2- i m_\rho \Gamma_{\rho}(q^2)},  
\end{equation}
where 
\begin{equation}
\Gamma_{\rho}(q^2)=\frac{g_\rho^2(q^2-4m_\pi^2)^{3/2}}{48\pi q^2}.
\end{equation}

Our last comment concerns the $\eta'\to\pi^+\pi^-\gamma$ decay amplitude, which can be easily found from (\ref{tot1}) by use of two obvious replacements  
\begin{eqnarray}
\label{tot2}
A_{\eta'\to\pi\pi\gamma}&=& \frac{e N_c}{12 \pi^{2} f_{\pi}^{3}} \frac{1}{\sqrt 3} \left( \frac{f_\pi}{f_8}\sin\theta_P+ \sqrt 2 \frac{f_\pi}{f_0} \cos\theta_P \right) \nonumber \\
&\times&\left[1 +\delta' +\frac{3q^2/a}{m_\rho^2-q^2- i m_\rho \Gamma_{\rho}(q^2)}\right] \nonumber \\
&\times& e_{\mu\nu\alpha\beta}\, \epsilon^{*\mu}(p_{\gamma})\, p_{\gamma}^\nu p_{+}^\alpha p_{-}^\beta ,
\end{eqnarray}
where $\delta'$ is obtained from $\delta$ by replacing $a_\eta\to a_{\eta'}$.

It is easy to verify (see, for instance, \cite{Holstein02}) that the expressions (\ref{tot1}) and (\ref{tot2}) differ from previously known estimates made on the basis of the VMD model only by the presence of $\delta$ and $\delta'$ terms -- contributions originated due to the difference between $\pi a_1$ and $\eta f_1$ mixing effects. Neglecting these terms ($\delta=\delta'=0$) we find
\begin{equation}
\Gamma_{\eta\to\pi\pi\gamma}^{\mbox{\scriptsize theor}}=63.08\,\mbox{eV}, \quad
\Gamma_{\eta'\to\pi\pi\gamma}^{\mbox{\scriptsize theor}}=64.06\,\mbox{keV}. 
\end{equation}
These results overestimate the experimentally measured partial widths
\begin{eqnarray}
&&\Gamma_{\eta\to\pi\pi\gamma}^{\mbox{\scriptsize exp}}=55.28\pm 3.2\,\mbox{eV},\quad \mbox{\cite{PDG18}}\\
&&\Gamma_{\eta'\to\pi\pi\gamma}^{\mbox{\scriptsize exp}}=58.60\pm 0.06\pm 1.08\,\mbox{keV}. \quad
\mbox{\cite{Ablikim19}}
\end{eqnarray}
The latter ones can be used to extract the values of
\begin{equation}
\label{deltas}
\delta=-0.1, \quad \mbox{and} \quad \delta'=-0.3.
\end{equation}

If one neglects $q^2$ terms in (\ref{tot1}) and (\ref{tot2}), one finds that factors at the corresponding kinematic parts of the amplitudes are $A^{(\prime)}= A_0^{(\prime)} (1+\delta^{(\prime)} )$, where the parameters for $\eta'$ decay are marked with a prime. These expressions are in agreement with the low-energy theorems \cite{Chanowitz75}. 

It is interesting also to note that similar parameters $\delta^{(\prime)}$ were considered in \cite{Meissner12} with the nearby estimates: $\delta =-0.22\pm 0.04$ and $\delta'=-0.40\pm 0.09$. A slight discrepancy with our results (\ref{deltas}) is apparently due to a more detailed account of unitary and analyticity corrections used in \cite{Meissner12} (see also \cite{Holstein98,Truong02}). 

In \cite{Meissner12} the analytical expression for $\delta$ has been established by the matching of the decay amplitude to the one-loop $U(3)$ extended ChPT result. The origin of our $\delta^{(\prime)}$ is associated with the procedure of elimination of the $PA$ mixing. They also can be calculated, for instance, in the extended NJL model. In this case they will depend on the angle of $f_1(1285) - f_1(1420)$ mixing, and apparently may be used to extract the numerical value of this angle from $\eta/\eta'\to\pi\pi\gamma$ decays.      

%%%%%%%%%%%%   FIG. 5   %%%%%%%%
\begin{figure}
\resizebox{0.50\textwidth}{!}{%
\includegraphics{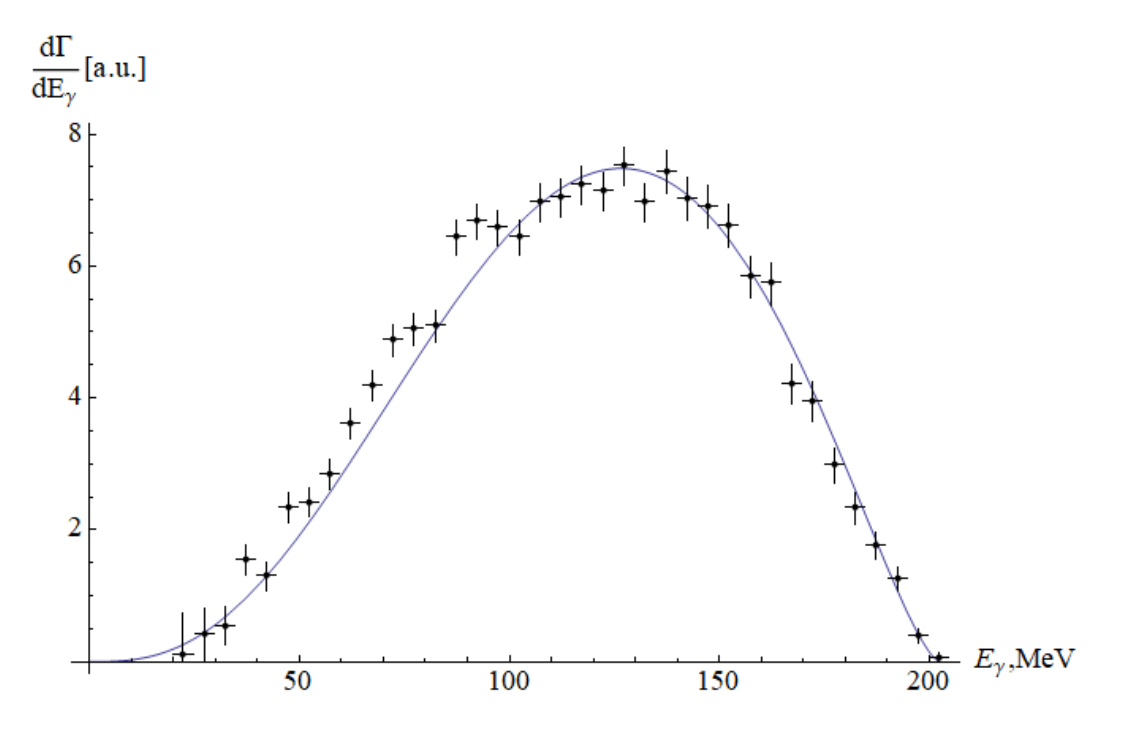}
}
\caption{The photon energy distribution $d\Gamma/dE_\gamma$ in the decay $\eta\to\pi^+\pi^-\gamma$. WASA-at-COSY data \cite{Adlarson12} are shown as crosses together with the result of the model fit for the width-modified amplitude (\ref{tot1}) and $\delta=-0.1$.}
\label{fig5}      
\end{figure}

%%%%%%%%%%%%   FIG. 6   %%%%%%%%
\begin{figure}
\resizebox{0.50\textwidth}{!}{%
\includegraphics{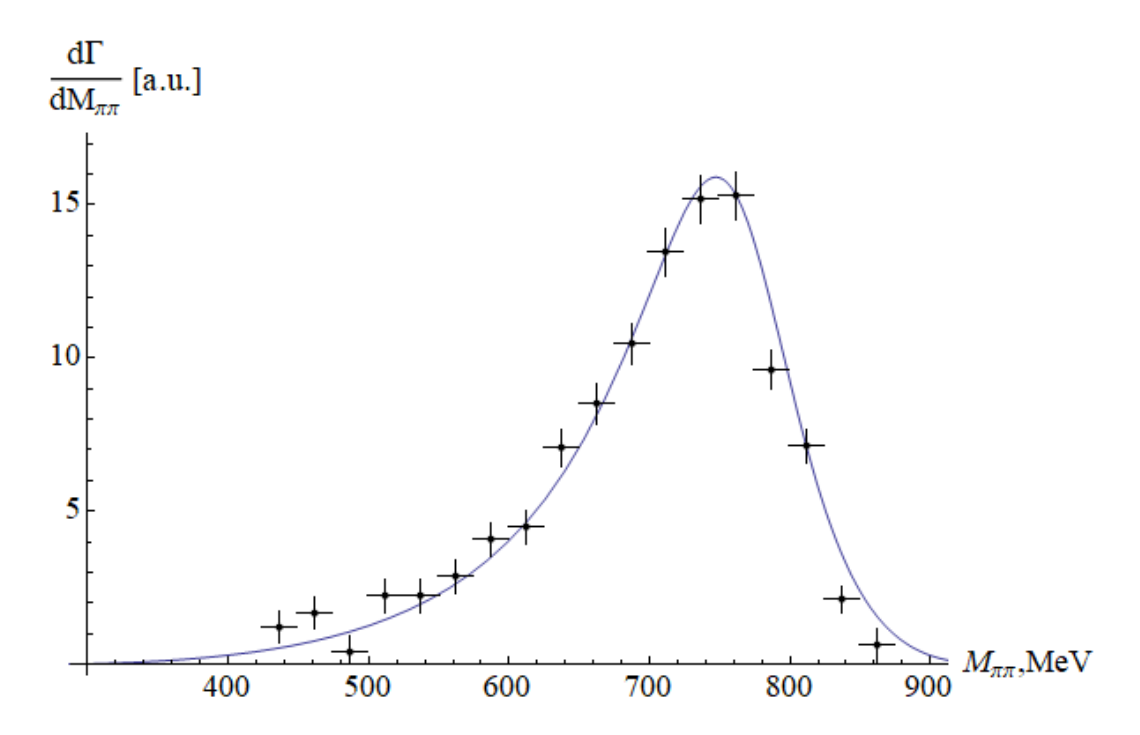}
}
\caption{$M_{\pi\pi}=\sqrt{q^2}$ distribution in the decay $\eta'\to\pi^+\pi^-\gamma$. Crystal Barrel data \cite{Abele97} are shown as crosses together with the result of the model fit for the width-modified amplitude (\ref{tot2}) and $\delta'=-0.3$.}
\label{fig6}      
\end{figure}

To conclude this section we confront the differential distributions for width-modified amplitudes (\ref{tot1}) and (\ref{tot2}) with the known experimental data. The found theoretical curves are shown in Fig.\ref{fig5} and Fig.\ref{fig6} compared to the experimental $E_\gamma$ spectrum of WASA-at-COSY collaboration \cite{Adlarson12} for $\eta$ case, and CRYSTAL BARREL data \cite{Abele97} for $\eta'$ case.   

\section{Conclusions}
The essential ambiguity related with surface terms of anomalous triangle diagrams has been used to satisfy the low-energy theorems for the $\eta/\eta'\to\pi^+\pi^-\gamma$ decays in the NJL model with spin-1 states. As a consequence, we have found that chiral anomaly not only determines the transition strength prefactor of these amplitudes $A_0^{(\prime)}$, but also explains the origin of the $U(3)$ breaking corrections accumulated in the non-resonant part described by the parameters $\delta^{(\prime)}$. The latter quantities have been extracted from the experiment. However their values are quite sensitive to the shape of the spectrum including the region where we do not have high quality data yet. The future more accurate data may significantly influence the integrated rate and therefore the values of $\delta^{(\prime)}$. For this reason, it seems reasonable to establish a solid framework for theoretical calculation of these parameters. The main result of our work is that it suggests a new important contribution for such calculations. We have shown that $\delta^{(\prime)}$ arise as a result of gauge covariant $PA$ diagonalization and are the residual $U(3)$ breaking effect after applying the Ward identities to the amplitudes of $\eta/\eta'\to\pi^+\pi^-\gamma$ decays. 

An important result of our work is also the fact that we were able to extend the known approach
\cite{Osipov18a,Osipov18b,Osipov20} to the description of more complex processes with $\eta, \eta'$ mesons. The non-trivial nature of the problem led to an interesting result -- an explanation of the appearance of the parameters $\delta^{(\prime)}$ in the amplitudes of these decays \cite{Meissner12}. 

In addition to the already mentioned applications of the result obtained here, we note the emerging new strategy for extracting the $1^{++}$ nonet singlet-octet mixing angle from the $\eta/\eta'\to\pi^+\pi^-\gamma$ decays. The extraction of $f_1(1285)-f_1(1420)$ mixing angle $\theta_{f_1}$ is associated with the processes directly related to the radiative decays of these mesons, or with the use of the Gell-Mann-Okubo mass formula together with the $K_1(1270)-K_1(1400)$ mixing angle \cite{Yang11}. It seems one can try to extract $\theta_{f_1}$ from the  $\eta/\eta'\to\pi^+\pi^-\gamma$ decays too. The reason is that the parameters $\delta^{(\prime)}$ most likely depend on this angle through the mechanism of $\eta, \eta' -f_1(1285), f_1(1420)$ mixings.

\begin{acknowledgements}
A.~A. Osipov and M.~M. Khalifa acknowledge financial support from FCT through the grant CERN/FIS-COM/0035/2019, and the networking support by the COST Action CA16201. The authors would like to thank B.~Hiller for her interest in the work and useful discussions.
\end{acknowledgements}

%\appendix
%\section{Appendix section}\label{app}

\end{document}